# Degree of entanglement for qubit-qutrit state

Safa Jami[1,2], Mohsen Sarbishei[1]

[1]Department of Physics, Ferdowsi University of Mashhad, 91775-1436 Mashhad, Iran.
[2]Department of Physics, Azad University of Mashhad, Mashhad, Iran.

*Abstract:* In this paper we consider a system consist of a qubit and a qutrit, and find a formula to evaluate the concurrence for it. We show that entanglement of formation for this system obeys the same relation as for two-qubits.

## I. Introduction:

One main goal of quantum theory is the characterization and quantification of the property of entanglement. Entanglement is the key resource in many of the recent quantum information applications like quantum teleportation [1], quantum dense coding [2] and cryptographic key distribution [3]. The importance of "entanglement as a resource" behooves us to quantify entanglement for as many quantum systems as possible. One of these systems is a system consist of a qubit and a qutrit, which we consider it and find a formula to evaluate the concurrence for it. We show that entanglement of formation for this system obeys the same relation as for two-qubits.

## II. Two-Qubits:

For the special case of bipartite pure state $\rho_{AB} = |\psi\rangle\langle\psi|$ a convenient measure of the degree of entanglement is the Von Neumann entropy of either of the two subsystems $A$ or $B$ [4]:

$$E|\psi\rangle = -tr(\rho_A \log_2 \rho_A) = -tr(\rho_B \log_2 \rho_B) \qquad (1)$$

Where $\rho_A (\rho_B)$ is the partial trace of $|\psi\rangle\langle\psi|$ over subsystem $A$ $(B)$. The measure $E(|\psi\rangle)$ is also known as the entanglement of formation (EOF) of $|\psi\rangle$ [5]. It can be shown that for the class of two qubits in the general pure state

$$|\psi\rangle = \sum_{i,j=1}^{2} a_{ij} |i,j\rangle \qquad (2)$$

Whit normalization $\sum_{i,j} |a_{ij}|^2 = 1$, the EOF is given by [5, 6]

$$E(|\psi\rangle) = h\left(\frac{1+\sqrt{1-C^2}}{2}\right) \qquad (3)$$

Where $h$ is the binary entropy function
$$h(x) = x \log_2 x - (1-x) \log_2(1-x) \qquad (4)$$
And where the concurrence $C$ is
$$C(|\psi\rangle_{2\times 2}) = 2|a_{00}a_{11} - a_{01}a_{10}| \qquad (5)$$
We also note that the concurrence (5) can be written equivalently as [7]
$$C(|\psi\rangle_{2\times 2}) = 2k_1 k_2 \qquad (6)$$

Where $k_1$ and $k_2$ are the two coefficients appearing in the Schmidt decomposition [8], $|\psi\rangle_{2\times 2} = k_1|x_1, y_1\rangle + k_2|x_2, y_2\rangle$ and $\{|x_1\rangle, |x_2\rangle\}$, $\{|y_1\rangle, |y_2\rangle\}$ are orthonormal bases for the Hilbert spaces of subsystems A and B, respectively.

### III. Qubit-Qutrit:

In this section we formulate a measure of entanglement for a pair of two and three level systems or qubit-qutrit in an arbitrary pure state. This measure is the generalization of the concurrence for two qubits defined in (5) or (6) to qubit-qutrit. We can write any state $\rho_{AB}$ of qubit-qutrit as

$$\rho_{AB} = \frac{1}{6}\left(I \otimes I + \vec{\sigma}^A \cdot \vec{u} \otimes I + \sqrt{3}\, I \otimes \vec{\lambda}^B \cdot \vec{v} + \sum_{i=1}^{3}\sum_{j=1}^{8} \beta_{ij} \sigma_i^A \otimes \lambda_j^B \right) \quad (7)$$

Where $\sigma_i$ $(i=1,2,3)$ are Pauli matrices and $\lambda_j$ $(j=1,\ldots,8)$ are the generators of SU(3), Gelmann matrices. Both $\vec{u} \equiv \{u_1, u_2, u_3\}$ and $\vec{v} \equiv \{v_1, \ldots, v_8\}$ can be regarded as vectors in a real, three and eight dimensional vector space, respectively. Eq. (7) follows from the fact that the reduced density matrices of subsystems A (qubit) and B are given by

$$\rho_A = tr_B \rho_{AB} = \frac{1}{2}\left(1 + \vec{\sigma}^A \cdot \vec{u}\right) \quad (8)$$

$$\rho_B = tr_A \rho_{AB} = \frac{1}{3}\left(1 + \sqrt{3}\,\vec{\lambda}^B \cdot \vec{v}\right) \quad (9)$$

The expansion coefficients in Eq. (7) are given by

$$u_i = tr(\rho_{AB} \sigma_i \otimes I) \quad (10)$$

$$v_j = \frac{\sqrt{3}}{2} tr(\rho_{AB} I \otimes \lambda_j) \quad (11)$$

$$\beta_{ij} = \frac{3}{2} tr(\rho_{AB} \sigma_i \otimes \lambda_j) \quad (12)$$

It is not difficult to show that if $\rho_{AB}$ corresponds to a pure state $\rho_{AB} = |\psi\rangle\langle\psi|$ then we have $|\vec{u}| = |\vec{v}|$. Furthermore if $|\psi\rangle$ is a product state then $|\vec{u}| = |\vec{v}| = 1$. Note on the other hand, that the case $|\vec{u}| = 0$ corresponds to the maximally mixed reduced density operators $\rho_A = \frac{1}{2}I$. This in turn implies that the original pure state $\rho_{AB} = |\psi\rangle\langle\psi|$ is maximally entangled.

Let's consider the quantity

$$C = \sqrt{1 - |\vec{u}|^2} = \sqrt{1 - |\vec{v}|^2} \quad (13)$$

It seems to be a good choice to measure the relative amount of entanglement contained in a pure state $\rho_{AB}$. Admittedly, the main motivation to choose C as a measure of the entanglement in a pure state of qubit-qutrit is that the concurrence for two-qubits can be equally defined by $C = \sqrt{1 - |\vec{u}|^2}$, where $\vec{u}$, the Bloch vector, is determining in (8). So we consider a general pure state of qubit-qutit of the form

$$|\psi\rangle_{2\times 3} = \sum_{i=0}^{1}\sum_{j=0}^{2} a_{ij}|i\rangle|j\rangle \qquad (14)$$

With normalization $\sum_{ij}|a_{ij}|^2 = 1$. For this state we evaluate $\vec{u}$ from (10), then the concurrence (13) is given by

$$C(|\psi\rangle) = 2\left[(a_{00}a_{11} - a_{01}a_{10})^2 + (a_{02}a_{10} - a_{00}a_{12})^2 + (a_{01}a_{12} - a_{02}a_{12})^2\right]^{\frac{1}{2}} \qquad (15)$$

This equation is the generalization of the concurrence for two qubits in (5) to qubit-qutrit.

The state in (14) can be written alternatively in terms of a Schmidt decomposition [8, 9]

$$|\psi\rangle_{2\times 3} = k_1|x_1 y_1\rangle + k_2|x_2 y_2\rangle \qquad (16)$$

Where $k_1$ and $k_2$ are real nonnegative coefficients satisfying $k_1^2 + k_2^2 = 1$ and $\{|x_1\rangle, |x_2\rangle\}$, $\{|y_1\rangle, |y_2\rangle\}$ are two orthonormal basis sets belonging to the Hilbert spaces of subsystems $A$ and $B$, respectively. Since $k_1$ and $k_2$ are unique for any given state $|\psi\rangle_{2\times 3}$ it should be possible to uniquely express $C(|\psi\rangle_{2\times 3})$ in terms of Schmidt coefficients $k_1$ and $k_2$. theory of the Schmidt decomposition [8], we know that $k_1^2$ and $k_2^2$ correspond to the two eigenvalues of $\rho_S = AA^\dagger$ or $\rho_B = A^\dagger A$ where $A$ is

$$A = \begin{pmatrix} a_{00} & a_{01} & a_{02} \\ a_{10} & a_{11} & a_{12} \\ 0 & 0 & 0 \end{pmatrix} \qquad (17)$$

These eigenvalues are solutions of the equation

$$\lambda^2 - \lambda + \frac{1}{4}C^2(|\psi\rangle_{2\times 3}) = 0 \qquad (18)$$

Whit $C(|\psi\rangle_{2\times 3})$ being the expression in Eq.(15). Denoting the roots of this equation by $k_1^2$ and $k_2^2$, they satisfy the relation

$$k_1^2 + k_2^2 = 1 \qquad (19)$$

$$k_1^2 k_2^2 = \frac{1}{4} C^2(|\psi\rangle_{2\times 3}) \qquad (20)$$

From the latter equation, we obtain the concurrence $C(|\psi\rangle_{2\times 3})$ in terms of the Schmidt coefficients associated whit $|\psi\rangle_{2\times 3}$

$$C(|\psi\rangle_{2\times 3}) = 2k_1 k_2 \qquad (21)$$

This as we expect, is the same as the concurrence for two qubits. This is obtained because for qubit-qutrit we have also two terms in Schmidt decomposition (16) like for two qubits. Since we can write the entanglement of formation in terms of Schmidt coefficients and as the relation between the concurrence and the Schmidt coefficients for qubit-qutrit is the same as for two qubits, one can conclude that the entanglement of formation for qubit-qutrit obeys the same relation as for two qubits

$$E(|\psi\rangle) = h\left(\frac{1 + \sqrt{1 - C^2}}{2}\right) \qquad (22)$$

Where $h$ is the binary entropy and $C$ is the concurrence for qubit-qutrit system.

In summary, in this paper we have presented a new measure to quantify the amount of entanglement for qubit-qutrit in a general pure state $|\psi\rangle_{2\times3}$. This measure has been expressed in two equivalent forms: either in terms of the expansion coefficients $a_{ij}$ of $|\psi\rangle_{2\times3}$ in an arbitrary basis $|ij\rangle$ (Eq. (15)); or else in terms of coefficients $k_i$ of $|\psi\rangle_{2\times3}$ in an arbitrary Schmidt basis $\{|x_i\rangle, |y_i\rangle\}$ (Eq. (21)). And finally we have derived the entanglement of formation for $|\psi\rangle_{2\times3}$.

―――――――――――